\documentclass[%
 aip,
 amsmath,amssymb,
 reprint,%
]{revtex4-1}

\usepackage{graphicx}
\usepackage{dcolumn}
\usepackage{bm}

\usepackage[utf8]{inputenc}
\usepackage[T1]{fontenc}
\usepackage{mathptmx}
\usepackage{etoolbox}
\usepackage[boxed]{algorithm2e}
\usepackage{color}

\usepackage[normalem]{ulem}

\makeatletter
\def\@email#1#2{%
 \endgroup
 \patchcmd{\titleblock@produce}
  {\frontmatter@RRAPformat}
  {\frontmatter@RRAPformat{\produce@RRAP{*#1\href{mailto:#2}{#2}}}\frontmatter@RRAPformat}
  {}{}
}%
\makeatother
\begin{document}

\preprint{AIP/123-QED}

\title{Accurate generation of stochastic dynamics based on multi-model Generative Adversarial Networks}

\author{Daniele Lanzoni}
\affiliation{Materials Science Department, University of Milano-Bicocca, Via R. Cozzi 55, I-20125 Milano, Italy}
\author{Olivier Pierre-Louis}
\affiliation{Institut Lumière Matière, UMR5306 Université Lyon 1—CNRS, 69622 Villeurbanne, France}
\author{Francesco Montalenti}
\affiliation{Materials Science Department, University of Milano-Bicocca, Via R. Cozzi 55, I-20125 Milano, Italy}

\date{\today}

\begin{abstract}

Generative Adversarial Networks (GANs) have shown immense potential in fields such as text and image generation. Only very recently attempts to exploit GANs to statistical-mechanics models have been reported. Here we quantitatively test this approach by applying it to a prototypical stochastic process on a lattice. By suitably adding noise to the original data we succeed in bringing both the Generator and the Discriminator loss functions close to their ideal value. Importantly, the discreteness of the model is retained despite the noise. As typical for adversarial approaches, oscillations around the convergence limit persist also at large epochs. This undermines model selection and the quality of the generated trajectories. We demonstrate that a simple multi-model procedure where stochastic trajectories are advanced at each step upon randomly selecting a Generator leads to a remarkable increase in accuracy. This is illustrated by quantitative analysis of both the predicted equilibrium probability distribution and of the escape-time distribution. Based on the reported findings, we believe that GANs are a promising tool to tackle complex statistical dynamics by machine learning techniques.

\end{abstract}

\maketitle

\section{Introduction}

Generative Adversarial Networks (GANs) \cite{goodfellow_generative_2014} are a class of Machine Learning (ML) methods capable of generating  data with the same statistical properties of an assigned training set. Importantly, this is accomplished without the need to explicitly estimate the target probability density, a daunting task for high-dimensional problems.

We recall that GANs are formulated~\cite{goodfellow_generative_2014} as an adversarial game between two neural networks (NNs), called Generator $G$ and Discriminator $D$, which are trained concurrently. 
The task of the Generator is to map samples $z$ extracted from a known probability distribution into samples resembling those drawn from the real (unknown) data distribution. These generated samples are passed to the Discriminator together with samples from the actual training dataset. The objective of the Discriminator is then to recognize whether its input comes from the generated or the true distribution. The two networks are trained in an alternate fashion by minimization of loss functions defining a zero-sum adversarial game. When Nash equilibrium is reached, in principle, $G$ successfully maps $z$ samples to elements which are distributed according to the true data distribution. 

In recent years, models exploiting GANs have shown impressive results in the generation of photo-realistic images \cite{karras_progressive_2017}, text \cite{yu_seqgan_2017} and in medical imaging \cite{yi_generative_2019}, only to cite a few applications in diverse fields.  On the other hand, successful applications of GANs in statistical mechanics/computational physics seems to be very limited when compared with other  ML approaches \cite{carleo_machine_2019}, 
which have been used, e.g., to map deterministic properties such as structure-energy relationships in molecular systems \cite{mueller_machine_2020, friederich_machine-learned_2021}, phase classification in materials \cite{chung_data-centric_2022} and prediction of complex nonlinear dynamical evolution~\cite{yang_self-supervised_2021, lanzoni_morphological_2022}. 
Nevertheless, GANs are a promising tool to model stochastic time series data \cite{brophy_generative_2023} and, specifically, stochastic dynamics in physical systems~\cite{yeo_generative_2022,stinis_sdyn-gans_2023}.

Despite their success, Generative Adversarial Networks are known to be particularly difficult to handle during training, as mode collapse and convergence failures are typically encountered~\cite{saxena_generative_2021}. Indeed, an intense activity on developing specialized architectures~\cite{radford_unsupervised_2015}, empirical "tricks"~\cite{goodfellow_nips_2017,saxena_generative_2021} and theoretical understanding~\cite{arjovsky_towards_2017, mescheder_numerics_2017} has been ongoing in last years. On the dynamical systems front, the recent Ref.~\onlinecite{yeo_generative_2022} focuses on GAN regularization techniques which allows the training of Recurrent Neural Networks capable of reproducing stochastic evolution of continuous physical systems. While alternative generative models such as diffusion denoising models~\cite{ho_denoising_2020} and variational autoencoders~\cite{kingma_auto-encoding_2022} are less prone to these convergence problems, the high quality of samples which can be achieved by GANs and their relatively fast generation times make them appealing.

In this work, we quantitatively test GANs by applying this approach to a prototypical stochastic process defined on a lattice. Convergence of the learning procedure close to Nash equilibrium and very high accuracy in the predicted dynamics is achieved upon exploiting two key procedures: noise injection and a simple, yet effective, multi-model average. These approaches are general and can be transferred to any GAN framework and architecture with straightforward modifications. While the stochastic process here considered is very simple, the reader should keep in mind that even in this one-dimensional settings, obtaining a high quality GAN can be non-trivial. On the other hand, the low dimensionality of the example, allows for an in-depth, quantitative analysis of the effect of the considered procedures. Additionally, this is the first example, to the authors knowledge, in which a GAN is employed to learn physical dynamics coming from lattice models, which are ubiquitous in computational physics. While the authors do not expect to obtain a more efficient method for generating stochastic trajectories in this specific case, we still hope that the current work will serve as a proof of concept and will stimulate further research in tackling more complex systems.

The paper is organized as follows. First a stochastic process on a lattice is defined, analytical expressions for equilibrium and kinetic properties of the system and the adaptation of the standard GAN approach are reported. Next, training convergence and the effects of noise injection as a regularization procedures are discussed. In particular, we show that this techniques greatly stabilize the training process and increases the quality of generated trajectories. This, however, is not enough to obtain accurate generation on the quantitative level. We therefore show that a suitable multi-model approach allows to recover quantitative agreement with close form expressions for both equilibrium and kinetic properties of the system. We conclude the Results section by outlining perspectives on transfer learning possibilities. For better clarity, we kept the most technical aspect of this work as appendices, while the main text reports results concerning optimal choices of hyperparameters.   

\section{Methods}

\subsection{Stochastic process definition}

\begin{figure}
    \centering
    \includegraphics[width=\columnwidth]{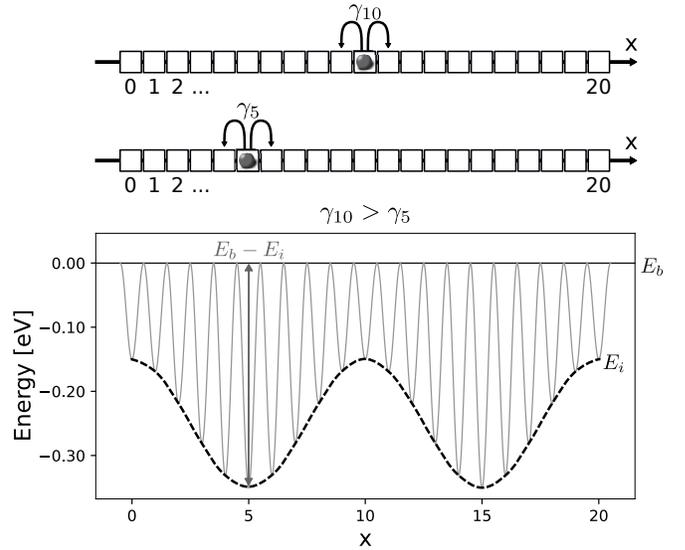}
    \caption{Graphical representation of the random walk. Lattice points correspond to energy minima. In order to reach a new lattice position, the particle has to overcome the diffusion barrier $E_b$. Gray curve represent the potential profile, dashed black curve interpolates $E_i$.}
    \label{fig::potential}
\end{figure}

In the diffusion process depicted on Fig.~\ref{fig::potential}, the rate at which a particle at the site $x_i$ moves to the right  and to the left  are identical and equal to  
\begin{align}
    \gamma_{i}=\nu{\mathrm e}^{-(E_b-E_i)/k_BT},
\end{align}
where $\nu$ is an attempt frequency, $E_i$ is the energy of state $i$, $E_b$ is the energy of the diffusion barrier, $k_B$ is Boltzmann's constant and $T$ is the temperature. For convenience, the lattice spacing is set to $1$. Positions take therefore integer values $x_i=i$. At domain boundaries, corresponding to $x=0$ ($x=L$) the particle is only allowed to jump to the right (left).

This diffusion process obeys detailed balance at equilibrium and the stationary probability of the Markov chain therefore corresponds to a well-defined equilibrium state with probability
\begin{align}
    P_{eq,i}=\frac{1}{\cal Z}{\rm e}^{-E_i/k_BT}
    \label{eq::equilibrium_distribution}
\end{align}
where ${\cal Z}=\sum_{i=1,L}\exp[-E_i/k_BT]$ is the partition function. 

We chose a two-well energy profile 
\begin{align}
    E_i=\frac{1}{2}(E_\text{max}+E_\text{min})
    +\frac{1}{2}(E_\text{max}-E_\text{min})\cos\left[\frac{4\pi i}{L}\right], 
\end{align}
where $L$ is the domain length. 

As a representative measure of the kinetic properties of the process, we use the average time $\tau$ for going from one minimum to the other. In the continuum limit $L\gg 1$~\cite{van_kampen_stochastic_1992}, 
\begin{align}
    \tau=\frac{Ld}{2}{\rm e}^{-(E_\text{max}+ E_\text{min})/2k_BT}
    I_0\left[\frac{E_\text{max}- E_\text{min}}{2k_BT}\right],
\end{align}
where 
$d$ is the distance between the minima of the potential, and $I_0$ is the modified Bessel function of the first kind.

In the following, we use $L=20$,
and eV's for the energy units, with $E_\text{min}=-0.35$, $E_\text{max}=-0.15$,  and $T=500$K. 

\subsection{Training set}

From the here-above defined stochastic process we generated  independent trajectories by Kinetic Monte Carlo (KMC) simulations, from which we extracted the elements of the training set, i.e. $(x_t,x_{t+\Delta t})$ pairs, where $x_t$ represents the particle position at time $t$ and $x_{t+\Delta t}$ the position at time $t + \Delta t$. 

Using time-units in which $\exp[E_b/k_BT]/\nu=1$, we set $\Delta t=10^3$, the mean residence times in sites with minimum and maximum energies being approximately $3 \Delta t$ and $\Delta t /30$. 
This specific choice of $\Delta t$ is not critical, except from some obvious requirements: $\Delta t$ much larger than the residence times would simply yield the long-time equilibrium distribution with no access to kinetic properties. At the other extreme, values of $\Delta t$ much smaller than residence times would call for too many iterations to generate meaningful trajectories. 

The full training set was built upon collecting $1200$ independent trajectories of $500$ snapshots $\Delta t$ apart from each other. The value of $t$ itself is a multiple of $\Delta t$, so that the training set is composed of couples $\{(x_0, x_{\Delta t}),(x_{\Delta t}, x_{2 \Delta t}),(x_{2\Delta t}, x_{3\Delta t}),...\}$. In the following, the term "step" will be used for such time intervals and not for individual Kinetic Monte Carlo steps, if not stated otherwise.

\subsection{GAN approach}

Let us now describe how GANs can be trained using this dataset. The task can be conveniently phrased in the framework of conditional GANs (cGANs) \cite{mirza_conditional_2014}. From a practical standpoint, the only difference with the original, unconditional GAN approach is that both the output of the Generator and that of the Discriminator are conditioned on some additional information. To mimic Markovian dynamics, the Generator is conditioned by the current state of the system. Hence, for assigned NN parameters $\theta_G$, to be optimized during training, the input of $G$ consists in the current position of the particle $x_t$ and in the latent random variable $z$. Hence, we write: 
\begin{equation}
    \tilde{x}_{t+\Delta t} = G(x_t, z|\theta_G),
\end{equation}
where $\tilde{x}_{t+\Delta t}$ is the generated next state of the system. The specific choice of the latent distribution is not crucial, provided that the dimension of $z$ is higher than the one of real data $x_{t+\Delta t}$~\cite{arjovsky_towards_2017}. In the present work, therefore, $z$ was conveniently sampled from a multivariate Gaussian distribution in $\mathbb{R}^{25}$ at each time step. The Generator produces real numbers and is initially unaware of both the confined and discrete characters of the stochastic dynamics
(i.e. $0\leq x_t\leq L$ and $x_t$ takes integer values).

Similarly, the scores of the Discriminator, indicating how likely is the observation of the next state $x_{t+\Delta t}$ are conditioned on $x_t$:
\begin{equation}
\begin{split}
    s = &D(x_{t+\Delta t}, x_t|\theta_D) \\
    \tilde{s} = & D(\tilde{x}_{t+\Delta t}, x_t|\theta_D) = D(G(x_t, z|\theta_G), x_t|\theta_D)
\end{split}
\end{equation}
where $s$ and $\tilde{s}$ are the scores for the next state in the dataset $x_{t+\Delta t}$ and of the generated next state $\tilde{x}_{t+\Delta t}$ respectively, while $\theta_D$ are the Discriminator parameters. As in the original GAN paper \cite{goodfellow_generative_2014}, scores are real numbers in $(0,1)$ and the last layer of the Discriminator comprises a sigmoid activation function, as is standard practice in NN classifiers~\cite{goodfellow_deep_2016}. Both the Generator and the Discriminator are implemented in the PyTorch framework \cite{paszke_pytorch_2019} as ResNet architectures \cite{he_deep_2015} with 7 hidden layers and have been trained using the Adam optimizer~\cite{kingma_adam_2014}.

Training proceeds by alternative minimization of the loss functions \cite{goodfellow_generative_2014}:
\begin{equation}
\label{eq::loss_functions}
    \begin{split}
        \mathcal{L}_G = & -\mathbb{E}[\log(\tilde{s})]\\
        \mathcal{L}_D = & -\frac{1}{2} \bigl(\mathbb{E} [\log(s)] + \mathbb{E}[ \log(1-\tilde{s})] \bigr)
    \end{split}
\end{equation}
where $\mathbb{E}$ is the expectation value operator. For the Generator, the non-saturating loss function discussed in Ref.~\onlinecite{goodfellow_generative_2014} has been used. At Nash equilibrium, both losses in Eq.~\ref{eq::loss_functions} are equal to $\log 2$. Pseudo-code and additional technical details on the training minimization procedure are reported in Appendix~\ref{app::training_porcedure}.

\begin{figure}
    \centering
    \includegraphics[width=\columnwidth]{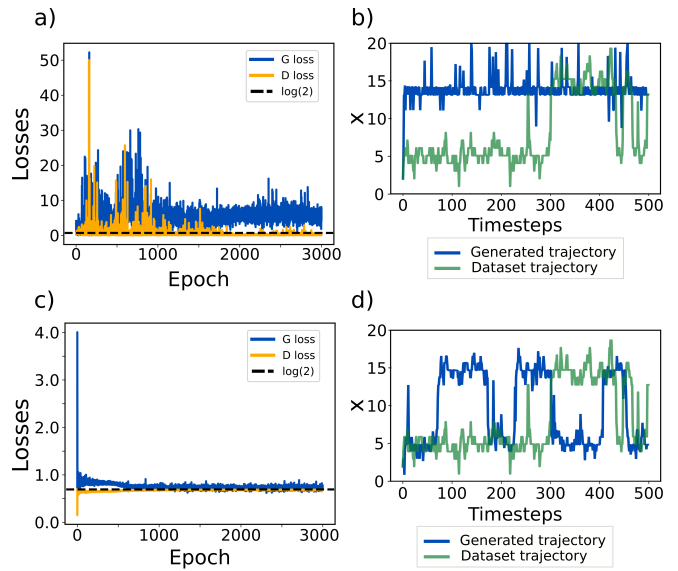}
    \caption{(a) $G$ and $D$ losses as a function of the number of epochs while minimizing equations \ref{eq::loss_functions} directly. Theoretical Nash equilibrium value of $\log 2$ is reported for reference. The behavior is clearly non-convergent. (b) Comparison of a generated trajectory (dark blue line) at the end of procedure in (a) and one from the dataset (transparent green line). (c) Lossplot obtained when Gaussian noise ($\sigma=0.25$) is added in the training procedure. (d) Example of a generated trajectory (dark blue line) obtained at the end of training procedure of (c) and one from the dataset (transparent green line).}
    \label{fig::lossplot_mixed}
\end{figure}

\section{Results}

\subsection{Noise injection and training convergence}

In Fig.\ref{fig::lossplot_mixed}a we report the behavior of the Generator and Discriminator loss functions during training. Results clearly show that  the training of the cGAN formulated above not only leads to oscillations, but to values of the loss functions far from the ideal $\log 2$ Nash equilibrium value. The Generator loss function exhibits strong oscillations, while $\mathcal{L}_D$ drops close to zero. This failure is also reflected in the poor quality of the generated trajectories, which can be directly assessed by visual inspection of the example reported in fig.~\ref{fig::lossplot_mixed}b. Noticeably, the trajectory exhibits a strong tendency to oscillate only near the minimum at $x=15$. This is reminiscent of mode collapse, one of the typical failure modes found in GANs~\cite{goodfellow_nips_2017, saxena_generative_2021}.

This unsatisfactory result can be traced back to a feature of the original formulation of the GAN approach which is frequently encountered \cite{saxena_generative_2021}. As pointed out by Arjovsky and Bottou \cite{arjovsky_towards_2017}, instabilities in training should be expected when the intersection between the support of the generated and the true data distribution has zero measure even in the continuous case. In our example, where true data live on a discrete lattice while $G$ outputs are real numbers, this problem is critical and results from Goodfellow et al.~\cite{goodfellow_generative_2014} are not expected to apply.
A possible solution also proposed in Ref.~\onlinecite{arjovsky_towards_2017} is based on adding random noise both to the true and  generated data before passing them to the Discriminator or the Generator. Similar strategies have also been suggested as a way to reduce extrapolation errors \cite{stinis_enforcing_2019} and are known to be equivalent to regularization \cite{bishop_training_1995}. For GANs training, this procedure expands of the support of probability distributions and solves the overlap issue. This explanation is supported by data reported in Appendix~\ref{app::discr_cont_comparison}, in which the role of additive noise was compared when training GANs on lattice and continuous dynamics.

In practice, Gaussian noise with zero mean and different standard deviations $\sigma$ were added to both true and generated examples:
\begin{equation}
    \label{eq::noise_addition}
    \begin{split}
        &x_\text{data} \rightarrow x_\text{data} + \varepsilon_\text{data} \\
        &G(x_\text{data}, z) \rightarrow G(x_\text{data}+\varepsilon_\text{data}, z) + \varepsilon_\text{gen}
    \end{split}
\end{equation}
where $x_\text{data}$ represent any instance of the particle position from the database. Random variables $\varepsilon_\text{data}$ and $\varepsilon_\text{gen}$ come from the same distribution, but are indicated with different symbols in order to stress their independence. Every time particle coordinates are passed to either the Generator or Discriminator, new random noise is added.

Increasing the value of $\sigma$ increases the speed of convergence and decreases
the fluctuations around Nash equilibrium. However, when $\sigma$ becomes too large,
the spatial distribution of the position convolved with the noise becomes smooth 
and the information associated
with the discreteness of the lattice positions is lost.  
In our case, $\sigma=0.25$ produced the best trajectories. 
A detailed analysis as a function of $\sigma$ is reported in Appendix~\ref{app::noise_strength_regularization}.
The loss plot for $\sigma=0.25$ is reported in figure \ref{fig::lossplot_mixed}c. The loss functions are now approaching the ideal value, despite some oscillations being still present (notice the smaller y-axis scale in~\ref{fig::lossplot_mixed}c as compared to~\ref{fig::lossplot_mixed}a).

During trajectory generation, no noise was added to the Generator output. Additive noise for $G$ inputs, however, has been retained, consistently with the training procedure. 
The quality of the generated trajectories is greatly increased with respect to the noiseless minimization, as seen in the example of fig.~\ref{fig::lossplot_mixed}d. Remarkably, despite the smoothing role of $\varepsilon$ in training, particle positions are closely peaked near discrete x values. Additionally, it is clear that the mode collapse problem encountered in the naive minimization approach has been strongly reduced. Both features has been further analyzed in Appendix~\ref{app::noise_strength_quality}.

\subsection{Multi-model generation of trajectories}

As already pointed out, oscillations of the loss functions of Fig. \ref{fig::lossplot_mixed}c are persistent.
This suggests that $G$s extracted at different epochs within the Nash equilibrium regime 
could exhibit significant differences.
We therefore compared equilibrium and kinetic properties of the trajectories produced by several Generators with those obtained by our reference KMC simulations. Equilibrium distributions from 4 different Generators extracted  within the Nash equilibrium regime and separated by 50 epochs from each other are reported in \ref{fig::equilibrium_distributions}a. They are obtained
from $10$ independent generated trajectories reaching a total time of $5 \times 10^5 \times \Delta t$.
Each single $G$ appears to produce qualitatively satisfactory
results. Indeed, the expected two-peak equilibrium distribution is obtained in all cases,
and as reported in Fig.~\ref{fig::lossplot_mixed} each Generator produces trajectories that looks reasonable.
However, significant quantitative variability is found in the distributions obtained by different $G$s. Furthermore,
each generated distribution is different from the true distribution 
extracted from KMC simulations, shown in Fig.\ref{fig::equilibrium_distributions}b.

Let us now show that this issue can be conveniently addressed by a multi-model scheme based on a suitable average which strongly increases the accuracy of predictions. This is done by randomly selecting Generators obtained at different epochs. Combining multiple models is often exploited in Machine Learning to increase their quality. Methods such as bagging \cite{breiman_bagging_1996}, boosting~\cite{freund_desicion-theoretic_1995} and stacking~\cite{wolpert_stacked_1992} represent only a handful of classical examples~\cite{ganaie_ensemble_2022}. In adversarial contexts, the use of model ensembles has been proposed during training in order to make models more resilient to attacks~\cite{tramer_ensemble_2020}.
The key issue in our case is that conditional probabilities are not explicitly accessible for a direct average. In order to circumvent this problem, we use the following procedure to generate a trajectory. Let us consider a set of models, each one associated with a different Generator extracted in the Nash equilibrium regime. At each time step $\Delta t$, we select randomly a model $j$ from the set composed by $M$ models. Invoking the law of total probability~\cite{ash_basic_1970}, the multi-model conditional probability $P_\text{mm}(x_{t+\Delta t}|x_t)$ is found to be equal to the average conditional probability:
\begin{equation}
\begin{split}
\label{eq::prob_average}
    P_\text{mm}(x_{t+\Delta t}|x_t) 
    &= \sum_j^M P_j(x_{t+\Delta t}|x_t) P(j) \\
    & =\frac{1}{M} \sum_j^M P_j(x_{t+\Delta t}|x_t)
\end{split}
\end{equation}
were $P_j(x_{t+\Delta t}|x_t)$ is the conditional probability for model $j$, $P(j) = 1/M$ is the (uniform) probability that model $j$ is chosen.

\subsection{Equilibrium distribution}

\begin{figure}
    \centering
    \includegraphics[width=\columnwidth]{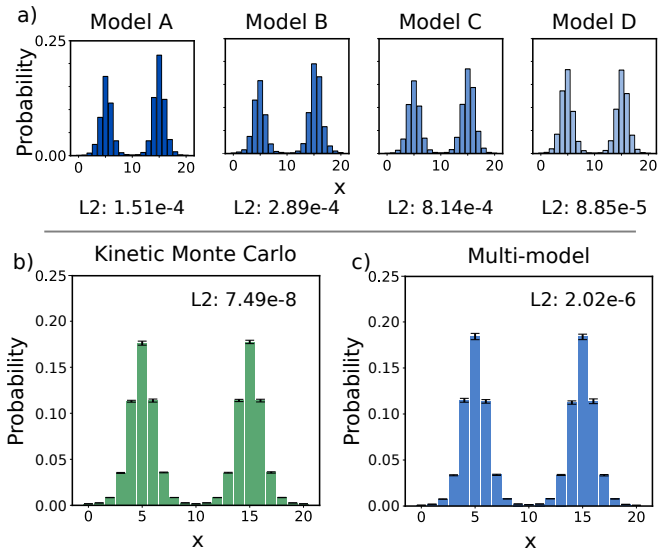}
    \caption{Equilibrium distribution obtained by random walk trajectories from different models. Models A, B, C and D in (a) refer to individual models picked at different epochs. (b) reports the equilibrium distribution obtained by Kinetic Monte Carlo simulations. (c) has been obtained by the multi-model approach. Black error bars represent confidence intervals (not reported in (a)). L2 distances between equilibrium distributions and the analytical one are also reported.}
    \label{fig::equilibrium_distributions}
\end{figure}

Figure~\ref{fig::equilibrium_distributions}c shows the equilibrium distribution as obtained from the aforementioned averaging procedure using $G$s at the end of the last 300 epochs of training.
Further discussions on the choice of the number of 
models and how they are extracted are reported in Appendix~\ref{app::G_number_choice}.
The distribution obtained in this way is remarkably close to that obtained by direct KMC simulations. Quantitatively, the L2 distance to the analytical equilibrium distribution drops by 2 orders of magnitudes, as reported in the insets of~\ref{fig::equilibrium_distributions}.

\subsection{Kinetic properties}

\begin{figure}
    \centering
    \includegraphics[width=0.8\columnwidth]{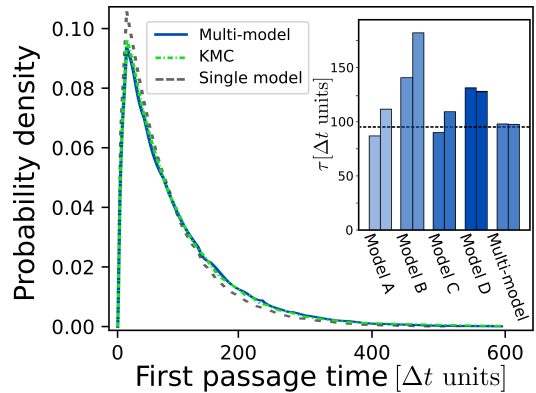}
    \caption{First passage time distribution obtained by generated trajectories (dark blue line), by KMC simulations (dashed-dotted green line) and by a Model A (dashed gray line). Inset reports first passage times as predicted by models at different epochs and by the multi-model approach. Models A, B, C and D are the same of figure~\ref{fig::equilibrium_distributions}. Black dashed line corresponds to the analytical mean value. Values for passages from the left (right) minimum to the right (left) one are reported separately.}
    \label{fig::hitting_times}
\end{figure}

Using the same multi-model procedure, we also generated $10^5$ independent trajectories and registered the time required to reach the left (right) minimum for a particle starting from the right (left) one. Distributions of first passage times from one minimum to the other, obtained from KMC simulations, the multi-model approach and one individual model, are reported in Fig.~\ref{fig::hitting_times}. KMC and multi-model distributions show remarkable quantitative agreement, confirming that the proposed method is indeed capable of providing an accurate description of kinetic properties of the system.

The inset of Fig.~\ref{fig::hitting_times} shows how the average first passage time $\tau$ changes with the multi-model procedure. Passage times for reaching the right minima from the left one and vice versa are reported separately. The theoretical expression in the continuum limit reported above predicts $\tau \approx 95.14$, while direct estimation from KMC simulations give $\tau \approx 95.2$. The multi-model approach recovers these values within a $3\%$ error.

Notice that the multi-model approach reported here
is more powerful than running every model in parallel and then taking average properties, since it is capable of generating more realistic individual trajectories, which allows for the analysis of kinetic pathways. We also remark that this approach is not computationally demanding, as all models are obtained during a single training procedure.

\subsection{Transfer learning perspectives}

Finally, we wish to emphasize that GANs could conveniently be used for transfer learning.
Indeed, as shown in Appendix~\ref{app::retraining}, a re-training of only a few epochs 
is sufficient to reach a new Nash equilibrium
when the diffusion potential is changed slightly. 
However, longer training might be necessary for important changes in the dynamics.

Moreover, the transient deviation of the loss during re-training,
which occurs only if the new samples have different statistical 
properties from the reference process,
could be used to discriminate the original stochastic process from
other stochastic processes. A few preliminary tests in this direction
are also reported in Appendix~\ref{app::retraining}. This opens novel directions to detect changes
as compared to a reference stochastic process. 

\section{Conclusions}

In conclusion, we have shown that Generative Adversarial Network architectures can be fruitfully applied to learn stochastic dynamics and are capable of reproducing both equilibrium and kinetic properties of a stochastic physical system. As compared previous attempts to learn stochastic processes based on GANs, our method does not rely on maximum mean discrepancy regularization~\cite{yeo_generative_2022,stinis_sdyn-gans_2023}. Furthermore, we have proposed a multi-model procedure which, coupled to simple noise-regularization, allows one to achieve quantitative learning of the target stochastic process. This procedure shares similarities with usual ensemble-learning strategies~\cite{tramer_ensemble_2020, Dong2020, Zhou2021}. However, our multiple models differ from usual ensemble learning strategies based on multiple training since they are extracted from a single learning run within the regime where the system fluctuates around Nash equilibrium. Additionally, our multi-model procedure differs in the sense that the average on the generators ensemble is taken as the system evolves in time, while in typical applications multiple models concur to a single prediction.

We have applied this procedure to the simple problem of diffusion in a potential, analyzing on quantitative grounds the outcomes of the learning procedure. While particular care is needed during training, we underline that the basic solutions presented here are expected to benefit from the versatility of Neural-Network based Machine-Learning methods
to address larger and more complex stochastic processes. In addition, GANs offer 
promising perspectives for transfer-learning and discrimination of stochastic processes.

\begin{acknowledgments}
F.M. acknowledges financial support from ICSC – Centro Nazionale di Ricerca in High Performance Computing, Big Data and Quantum Computing, funded by European Union – NextGenerationEU”.
\end{acknowledgments}

\section*{Data Availability Statement}

The data that support the findings of this study are available
from the corresponding authors upon reasonable request.

\appendix

\section{Training Procedure}
\label{app::training_porcedure}

In this section, we provide more details on the training procedure employed for the minimization of the adversarial Loss functions:

\begin{equation}
\label{eq::loss_functions_app}
    \begin{split}
        \mathcal{L}_G = & -\mathbb{E}[\log(\tilde{s})]\\
        \mathcal{L}_D = & -\frac{1}{2} \bigl(\mathbb{E} [\log(s)] 
        + \mathbb{E} [\log(1-\tilde{s})] \bigr)
    \end{split}
\end{equation}
where $\mathcal{L}_G$ and $\mathcal{L}_D$ are the Generator and Discriminator loss functions respectively, $s$ and $\tilde{s}$ are the scores provided by the Discriminator and $\mathbb{E}$ represent the expectation value operator.
Minimization has been performed using standard PyTorch implementation \cite{paszke_pytorch_2019} of the Adam optimizer \cite{kingma_adam_2014}, with batches composed of $\approx 7000$ coordinates pairs. Adam parameters have been set to $\beta_1 = 0.5$, $\beta_2 = 0.999$ and learning rate was $10^{-4}$. Pseudocode for the optimization is reported in Algorithm~\ref{alg::pseudocode}. As the Discriminator should be at optimality for a fixed Generator \cite{goodfellow_generative_2014}, 20 $\mathcal{L}_D$ minimization steps were performed for every $\mathcal{L}_G$ minimization step. In our tests, this produced the best trade off between quality of the generated trajectories and training time.

\begin{algorithm}
\label{alg::pseudocode}
\caption{Algorithm used to minimize eq.~\ref{eq::loss_functions_app}. \textbf{for} loop iterating on the dataset operates on batches. ${\cal N}(0, I_{25})$ indicates a multivariate normal distribution in $\mathbb{R}^{25}$ having the identity as covariance matrix, while ${\cal N}(0, \sigma)$ represents the standard normal distribution with standard deviation $\sigma$.}

$tot\_epochs \gets \text{set number of epochs}$ \;
$\sigma \gets \text{set additive noise standard deviation}$ \;
$G \gets \text{initialize Generator}$ \;
$D \gets \text{initialize Discriminator}$ \;

\For{epoch in tot\_epochs }
{
    \For{ $(x_t, x_{t+\Delta t})$ in dataset }
    {
        $(\varepsilon_1, \varepsilon_2) \gets \text{samples from } {\cal N}(0,\sigma)$ \;
        $x_t \gets x_t + \varepsilon_1$ \;
        $x_{t+\Delta t} \gets x_{t+\Delta t} + \varepsilon_2$ \;

        \For{20 iterations}
        {
            $z \gets \text{sample from } {\cal N}(0,I_{25})$ \;
            $\varepsilon_3 \gets \text{sample from } {\cal N}(0, \sigma)$\;
            $\tilde{x}_{t+\Delta t} \gets G(x_t, z)$ \;
            $\tilde{x}_{t+\Delta t} \gets \tilde{x}_{t+\Delta t} + \varepsilon_3$ \;
            $\tilde{s} \gets D(\tilde{x}_{t+\Delta t}, x_t)$ \;
            $s \gets D(x_{t+\Delta t, x_t})$ \;
            $\mathcal{L}_D \gets -\frac{1}{2} [ log(s) + log(1-\tilde{s}) ]$ \;
            $\text{Adam update for } \theta_D$ \;
        }

        $\tilde{s} \gets D(\tilde{x}_{t+\Delta t}, x_t)$ \;
        $\mathcal{L}_G \gets -log(\tilde{s}) $ \;
        $\text{Adam update for } \theta_G$\;
    
    }

    $\text{Save } G \text{ and } D$
}
\end{algorithm}

\section{Noise regularization and discrete stochastic dynamics}
\label{app::discr_cont_comparison}

The stabilization effect of additive noise during training is a critical step whenever the stochastic dynamics to be learned lives on a discrete lattice. This statement follows Goodfellow et al.~\cite{goodfellow_deep_2016}, who claimed in their original paper that "GANs require differentiation through the visible units, and thus cannot model discrete data". This is also consistent with the work of Arjovsky and Bottou~\cite{arjovsky_towards_2017}, which shows that GANs training stability depends on the intersection conditions of the true data and generated probability distribution supports.

As a test, we trained GANs on two prototypical stochastic processes: a Gaussian step and a discrete step one-dimensional random walk without any external potential. In both cases, a particle performs an unbiased one-dimensional random walk on the interval $[0,20]$. In the first model, at every timestep, a standard normal random variable is added to the position of the particle. In the second model, the particle takes instead a unit step to the left or to the right with equal probabilities. Motion is confined in $[0,20]$ by rejecting moves which would place the particle outside such interval. Datasets for both the continuous and the discrete model were constructed, each comprising $(x_t, x_{t+\Delta t})$ pairs coming from $200$ independent trajectories composed by $10^4$ random walk steps. In order to mimic the procedure reported in the main text, in which the time interval between generated states can be bigger than individual diffusion steps, pair elements were spaced by a $\Delta t$ corresponding to $10$ KMC steps.

\begin{figure}
    \centering
    \includegraphics[width=\columnwidth]{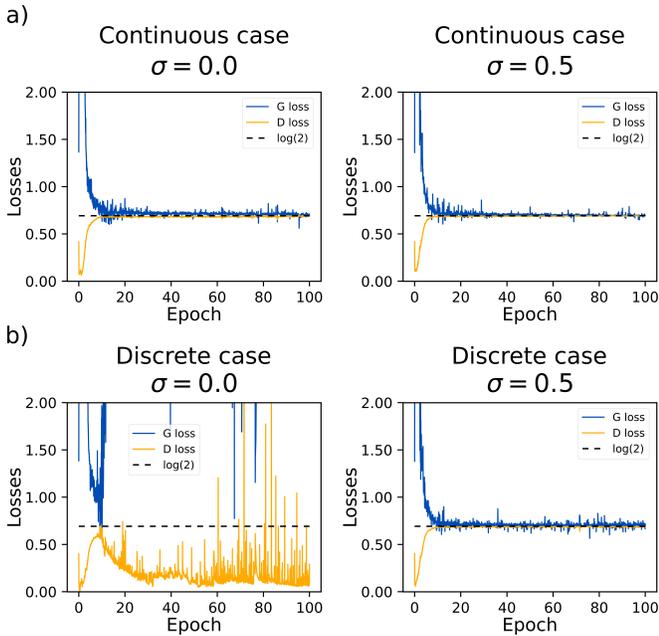}
    \caption{(a) Lossplot obtained by training a GAN on trajectories obtained by the random walk on continuous values. (b) Reports the same loss functions in the case of training on discrete dynamics.}
    \label{fig::gaussian_integer_walk_loss}
\end{figure}

Fig.~\ref{fig::gaussian_integer_walk_loss}a reports the training loss function from a $G$-$D$ couple with the same hyper-parameters as in the main text trained on $(x_t, x_{t+\Delta t})$ coming from the Gaussian walk model. While noise regularization slightly reduces oscillations, it does not seem to be critical for the continuous case, as both loss functions rapidly converge towards the theoretical $\log 2$ value.
On the other hand, the $\sigma=0.0$ case for the discrete dynamics shows a clearly non-convergent behavior (Fig.~\ref{fig::gaussian_integer_walk_loss}b). This is readily solved when additive noise is introduced in the training procedure. 

The noise-regularization procedure, therefore, seems to be critical for model convergence when the GAN approach used in the present work is applied to discrete stochastic systems. In continuous systems, on the other hand, it plays a less crucial role but still allows for a reduction in loss functions oscillations around the Nash value $\log 2$.

\section{Training results as a function of noise strength}
\label{app::noise_strength_regularization}

\begin{figure}
    \centering
    \includegraphics[width=\columnwidth]{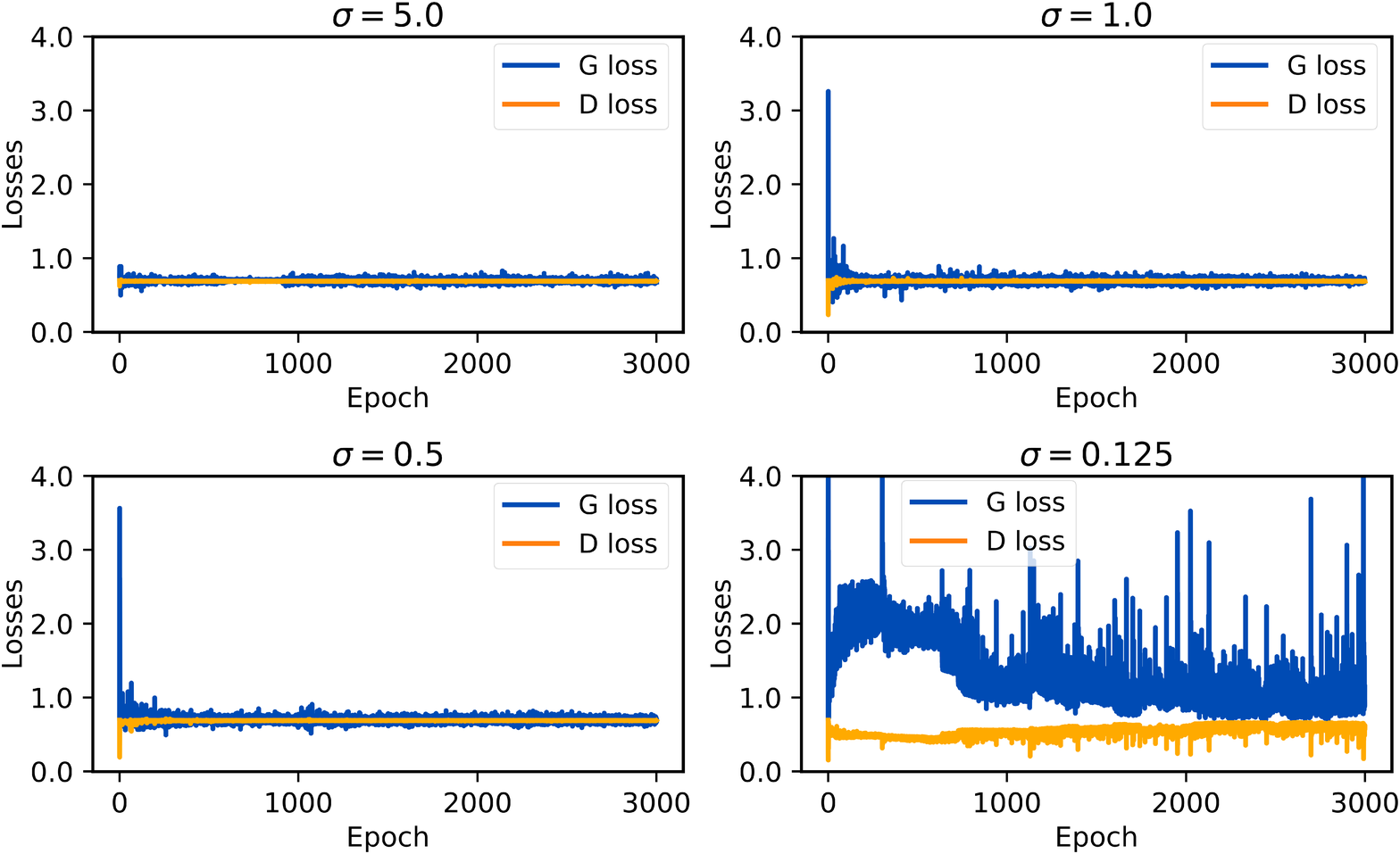}
    \caption{Lossplot for different values of the training noise $\sigma$. The lower the standard deviation of the additive noise, the lower is the stabilization effect.}
    \label{fig::lossplot_supp}
\end{figure}

In the main text, we discussed results for a noise standard deviation of $\sigma = 0.25$, claiming that, among tested values, it provides the best results in terms of training convergence speed and quality of the generated trajectories. We report in Fig.~\ref{fig::lossplot_supp} the lossplots for other values of $\sigma$.

As the value of $\sigma$ decreases, the oscillations in loss functions during training become stronger. In the $\sigma=0.125$ case oscillations are so strong that the $3000$ epochs used here are not sufficient to conclude if convergence can be reached.

These tests show that higher variance for $\varepsilon$ leads to more stable training procedures. Notice, however, that this does not suffice for $\sigma$ selection, as lossplots do not contain quantitative information on the quality of the generated trajectories.

\section{Quality of generated trajectories 
as a function of noise strength}
\label{app::noise_strength_quality}

\begin{figure}
    \centering
    \includegraphics[width=\columnwidth]{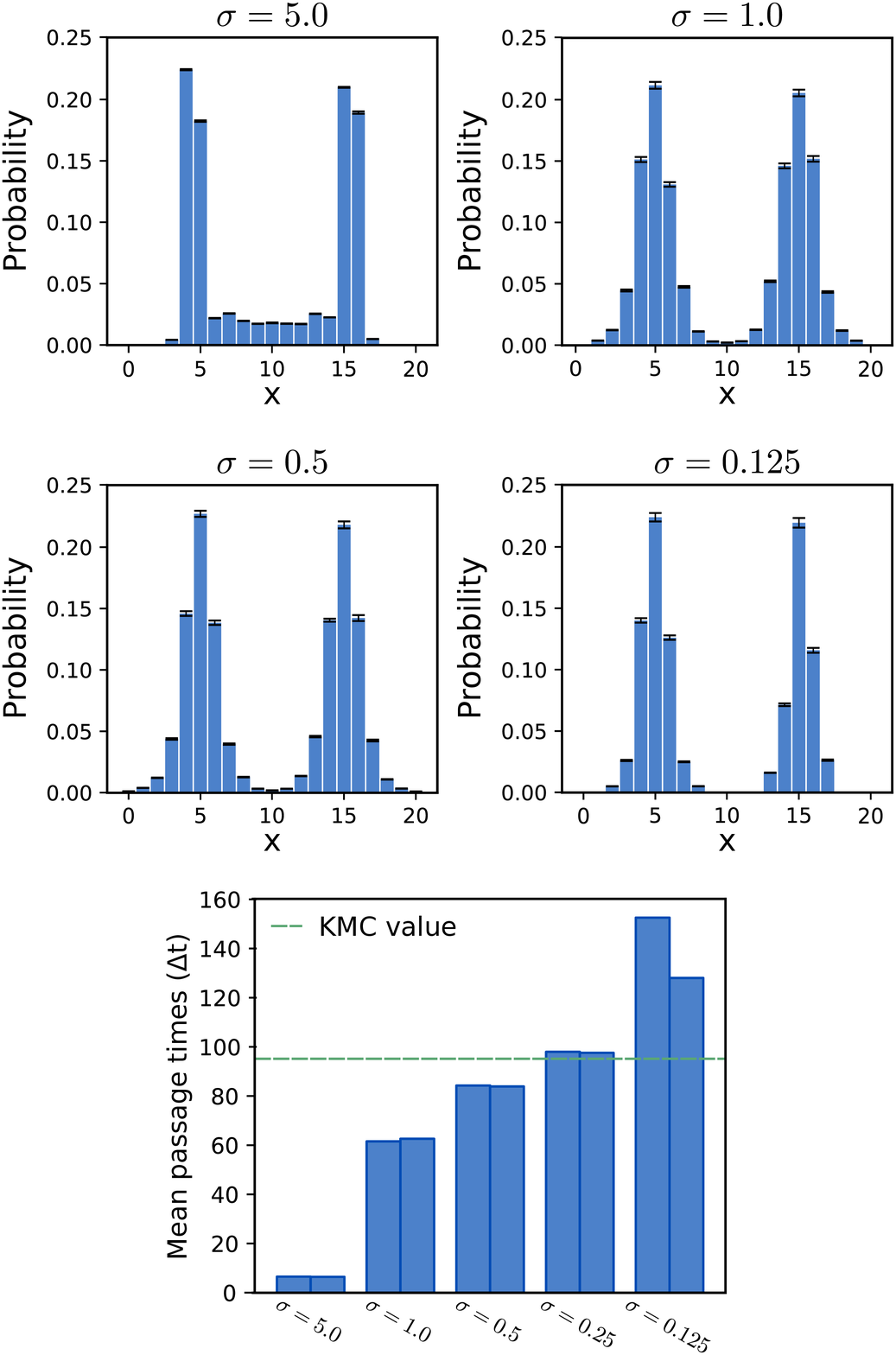}
    \caption{Equilibrium distributions obtained by Generators trained with different additive noise strength. Black bars represent confidence interval for probabilities. Barplot on the bottom reports mean passage times. For every $\sigma$ value, the two bars represent values for left to right and right to left passages respectively. KMC estimate of mean passage time is reported as a green dashed line.}
    \label{fig::equilibrium_supp}
\end{figure}

We now report an analysis of the trajectories produced by training GANs with non-optimal values of  $\sigma$. Fig.~\ref{fig::equilibrium_supp} shows the equilibrium distributions obtained for the same values as those used in the previous section. These distributions have been obtained through the same procedure as that reported for $\sigma=0.25$ in the main text, using the multi-model approach.

It can be noticed that at high $\sigma$ values details on the equilibrium distribution are lost. For example, for $\sigma=5.0$ the Generator is merely capable of recovering the most basic aspect of the dynamics, i.e. the presence of two peaks in the equilibrium distribution. In contrast, for $\sigma=0.125$, the probability density is suppressed in regions close to potential energy maxima, i.e. configurations that are rarely visited.

These learning failures can also be observed in the mean passage times, reported in the bottom panel of Fig.~\ref{fig::equilibrium_supp}, where the values for $\sigma=0.25$ are also reported for reference. Strong noise leads to a spurious reduction in learned passage times, and $\sigma=0.125$, on the other hand, exhibits significantly longer $\tau$.

\begin{figure}
    \centering
    \includegraphics[width=\columnwidth]{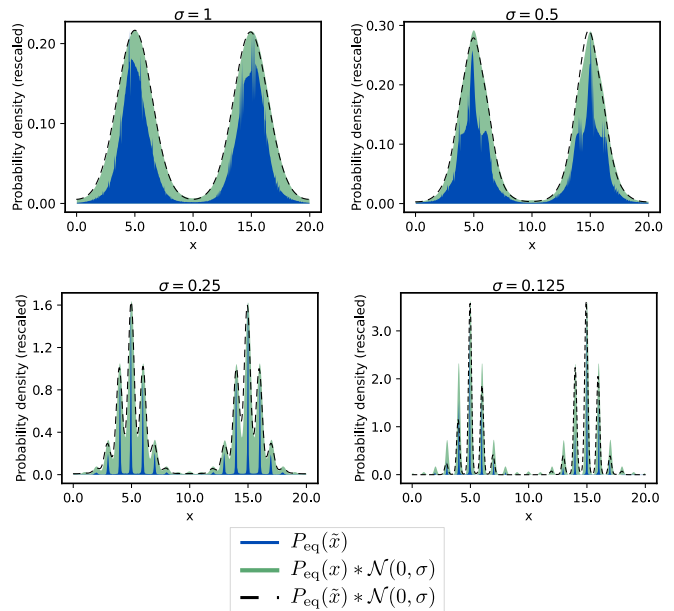}
    \caption{Equilibrium distributions obtained by KMC convolved with the noise distribution ($P_\text{eq}(x)*\mathcal{N}(0,\sigma)$, pale green), and by generators, both with ($P_\text{eq}(\tilde{x})*\mathcal{N}(0,\sigma)$, dark blue) and without convolution ($P_\text{eq}(\tilde{x})$, dashed black line).}
    \label{fig::sigma_conv}
\end{figure}

Noise strength has an additional effect on the generated trajectories: the lower the noise standard deviation, the more discrete-like the trajectories appear. This property can be appreciated in Fig.~\ref{fig::sigma_conv}, where the generated equilibrium distributions $P_\text{eq}$
as in \ref{fig::equilibrium_supp} are plotted in dark blue with smaller discretization bins ($0.04$ in the units of lattice spacing). Probability densities are rescaled for clarity. For $\sigma=5.0$ (not reported), $\sigma=1.0$ and $\sigma=0.5$ the learned probability density is spread out into two broad peaks, corresponding to the two minima in the potential energy function. On the other hand, the two cases with $\sigma=0.25$ and $\sigma=0.125$ exhibit a comb-like distribution, with sharp peaks at each lattice site.

This could be explained by considering how the additive noise affects the Nash equilibrium condition for GAN training. Following Ref.~\onlinecite{arjovsky_towards_2017}, the Generator task is now to produce samples $\tilde{x}_{t+\Delta t}$ such that, once noise $\varepsilon$ is added, are distributed as $x_{t+\Delta t}+\varepsilon$. Close to Nash equilibrium, therefore, Generator and true data (conditional) probability densities, $P_\text{g}$ and $P_\text{data}$, should be related via
\begin{equation}
    \label{eq::modified_nash}
    P_\text{g}(\tilde{x}_{t+\Delta t}|x_t)*\mathcal{N}(0,\sigma) \approx P_\text{data}(x_{t+\Delta t}|x_t)*\mathcal{N}(0,\sigma),
\end{equation}
where $*$ denotes the standard convolution product and $\mathcal{N}(\mu, \sigma)$ a normal distribution with mean $\mu$ and standard deviation $\sigma$. 

In Fig.~\ref{fig::sigma_conv}, equilibrium distributions convolved with the respective $\mathcal{N}( 0 ,\sigma)$ are reported (pale green). Clearly, information about the lattice discreteness
is completely lost for large $\sigma$, which prevents learning microscopic details on valid particle positions.
However, the training converges to the modified Nash equilibrium of eq.~\ref{eq::modified_nash}, as confirmed by the observation that
the convolution of the generated equilibrium distribution $P_\text{eq}(\tilde{x})*\mathcal{N}(0,\sigma)$ represented by the dashed black lines in~\ref{fig::sigma_conv} is remarkably close to the target one $P_{eq}(x)*\mathcal{N}(0,\sigma)$. We also remark that for $\sigma=0.25$, despite the sharp nature of the peaks in the probability distribution, there is no significant mode collapse, as all lattice position are visited. On the contrary, for other values of $\sigma$, a depletion of probability density in energy-maxima regions can be clearly observed.
In summary, these results show how poor trajectories and good training convergence can coexist. One therefore should not rely solely on $\mathcal{L}_G$ and $\mathcal{L}_D$ values.

This analysis also shows how the optimal value of $\sigma$ comes from the trade off between training stabilization and the loss of microscopic details of the dynamics. Additionally, it provides a physics-based criterion to choose noise strength, as it is closely related to the distance between accessible states in stochastic dynamics on discrete lattices.

\section{Choice of the number of Generators in the multi-model approach}
\label{app::G_number_choice}

In this section we provide more details on the effect of the number of Generators used in the multi-model approach. In the main text, NN models at the end of the last $300$ training epochs were chosen, as the simple architecture considered in this work does not pose memory or computational constraints.

\begin{figure}
    \centering
    \includegraphics[width=\columnwidth]{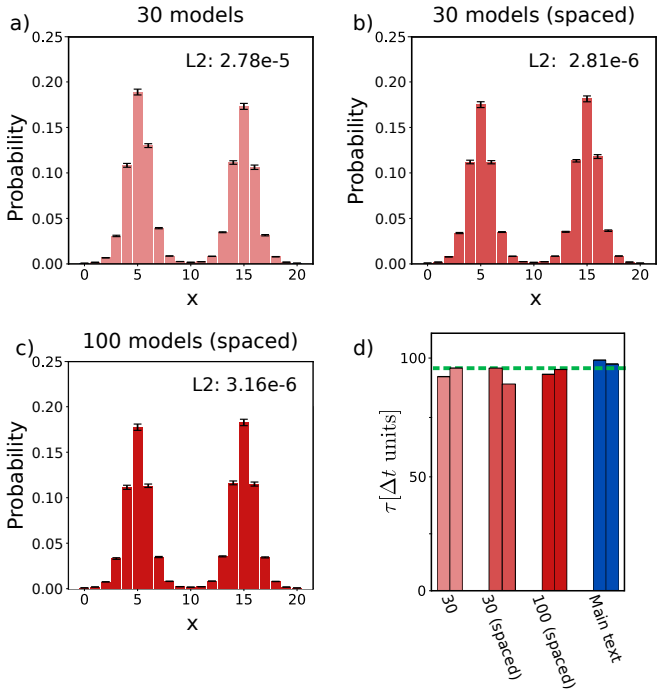}
    \caption{Results obtained through the multi-model approach by using different models for the average. (a), (b) and (c) represent the equilibrium distribution obtained using Generators at the end of the last $30$ epochs, by selecting $1$ Generator every $10$ in the last $300$ epochs and $100$ models spaced by $10$ epochs in the last $1000$. (d) reports mean first passage times (main text bars are reported for comparison; left to right and right to left passages are reported separately). KMC value is reported as a dashed green line.}
    \label{fig::multi_model_alternatives}
\end{figure}

Fig.~\ref{fig::multi_model_alternatives}, shows equilibrium distribution and mean passage times
with different ways to choose the Generators for the multi-model method: the last $30$ Generators in the training, $30$ Generators spaced by $10$ epochs in the last $300$ and $100$ Generators spaced by $10$ epochs in the last $1000$. The first choice (Fig.~\ref{fig::multi_model_alternatives}a) leads to worst quality in the generated trajectories,
but is still an improvement with respect to the single model prediction.
The $30$ spaced models (Fig.~\ref{fig::multi_model_alternatives}b) performs very similarly to last $300$ used in the main text, and provides results that are also similar to $100$ spaced Generators in terms of L2 distance with the analytical distribution.
Moreover, as seen from \ref{fig::multi_model_alternatives}, there is no significant changes in the accuracy of the first passage times with our different numbers of models. In summary, the best tradeoff in terms of computational cost and accuracy among those that we have tested is to use 
30 Generators spaced by 10 epochs, while the $300$ models choice reported in the main text produces slightly more accurate results.

\section{Re-training for transfer learning and discrimination}
\label{app::retraining}

\begin{figure}
    \centering
    \includegraphics[width=\columnwidth]{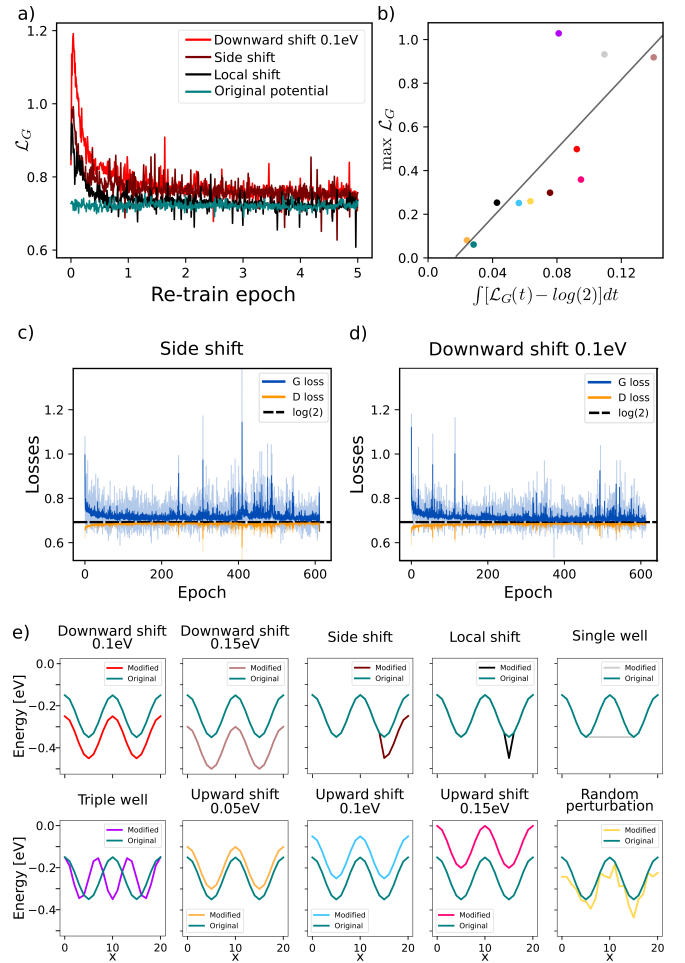}
    \caption{ (a) Loss-plot for re-training with different potentials. Perturbations correspond to a downward shift of the potential by 0.1eV for all points ("Downward shift 0.1eV"), for points at $x \ge 15$ ("Side shift") and for the point a t $x=15$ ("Local shift"). The teal curve "Original potential" has been obtained from new trajectories coming from the original potential of fig.~1 of the main text. (b) Correlation plot between the peak height and the integral of the loss function over five epochs. Pearson correlation coefficient is $\approx 0.801$. The points correspond to the 10 potentials reported in (e), and the green point corresponds to re-training with a new data set produced with the initial potential. (c,d) Longer re-trainings.  (e) Perturbed potentials.}
    \label{fig::detection}
\end{figure}

In this section, we propose to re-train GANs 
with a new KMC trajectory dataset after a converged
training on the original dataset. The new dataset is produced
by diffusion in a modified potential. Our aim is to investigate
possible perspectives of using GANs for transfer-learning
or for discrimination between the original dataset and datasets produced by
other stochastic processes.

The re-training loss $\mathcal{L}_G$ is reported in Fig.~\ref{fig::detection}a for four different datasets.  Perturbations of the potential were performed by a downward shift of the potential by $0.1eV$, respectively for the whole potential ("Downward shift 0.1eV" in \ref{fig::detection}(e)), for lattice points at $x \ge 15$ ("Side shift" in \ref{fig::detection}(e)) and for the single point at $x=15$ ("Local shift" in \ref{fig::detection}(e)). An additional curve for new trajectories coming from the same potential ("Original potential") is also reported. For stronger perturbations in the potential, the initial peak in the loss function and the area between $\mathcal{L}_G$ and $\log 2$ are larger. While the loss relaxes to the Nash equilibrium value after a few epochs for the local shift, the relaxation is slower for larger changes of the potential ("Global shift" and "Side shift"). \ref{fig::detection}b,c reports the losses obtained during a longer training. The relaxation towards Nash equilibrium is faster than the initial training (see Fig.3 of the main text), but is of the same order of magnitude. These results suggest that small perturbations of the potential can be learnt quickly and open interesting directions for future research on transfer learning.

Re-training also opens perspectives in the discrimination between different stochastic processes. Indeed, as a byproduct of the training procedure, the Discriminator should in principle contain implicit information on the distribution of the Generator and true data. However, near convergence the output of the Discriminator should be close to $1/2$. Therefore, we have not been able to use it directly to identify whether a new dataset comes from dynamics on the same potential or has been produced by a different stochastic process. However, since it leads to a deviation of the losses from the Nash equilibrium value, re-training does distinguish between different datasets, as shown in \ref{fig::detection}. Furthermore, the deviation of the losses occurs already in the first epochs of retraining. Hence, distinguishing between the original dataset and other datasets can be done quickly even if retraining has not converged to Nash equilibrium. As an additional supporting data for the fact that the initial peak of the loss contains robust information about the full relaxation during longer retraining, \ref{fig::detection}b shows that the peak value is correlated with the integrated deviation of the loss from the Nash equilibrium value over 5 epochs.

\nocite{*}
\bibliography{biblio}

\begin{thebibliography}{38}%
\makeatletter
\providecommand \@ifxundefined [1]{%
 \@ifx{#1\undefined}
}%
\providecommand \@ifnum [1]{%
 \ifnum #1\expandafter \@firstoftwo
 \else \expandafter \@secondoftwo
 \fi
}%
\providecommand \@ifx [1]{%
 \ifx #1\expandafter \@firstoftwo
 \else \expandafter \@secondoftwo
 \fi
}%
\providecommand \natexlab [1]{#1}%
\providecommand \enquote  [1]{``#1''}%
\providecommand \bibnamefont  [1]{#1}%
\providecommand \bibfnamefont [1]{#1}%
\providecommand \citenamefont [1]{#1}%
\providecommand \href@noop [0]{\@secondoftwo}%
\providecommand \href [0]{\begingroup \@sanitize@url \@href}%
\providecommand \@href[1]{\@@startlink{#1}\@@href}%
\providecommand \@@href[1]{\endgroup#1\@@endlink}%
\providecommand \@sanitize@url [0]{\catcode `\\12\catcode `\$12\catcode
  `\&12\catcode `\#12\catcode `\^12\catcode `\_12\catcode `\%12\relax}%
\providecommand \@@startlink[1]{}%
\providecommand \@@endlink[0]{}%
\providecommand \url  [0]{\begingroup\@sanitize@url \@url }%
\providecommand \@url [1]{\endgroup\@href {#1}{\urlprefix }}%
\providecommand \urlprefix  [0]{URL }%
\providecommand \Eprint [0]{\href }%
\providecommand \doibase [0]{http://dx.doi.org/}%
\providecommand \selectlanguage [0]{\@gobble}%
\providecommand \bibinfo  [0]{\@secondoftwo}%
\providecommand \bibfield  [0]{\@secondoftwo}%
\providecommand \translation [1]{[#1]}%
\providecommand \BibitemOpen [0]{}%
\providecommand \bibitemStop [0]{}%
\providecommand \bibitemNoStop [0]{.\EOS\space}%
\providecommand \EOS [0]{\spacefactor3000\relax}%
\providecommand \BibitemShut  [1]{\csname bibitem#1\endcsname}%
\let\auto@bib@innerbib\@empty
\bibitem [{\citenamefont {Goodfellow}\ \emph {et~al.}()\citenamefont
  {Goodfellow}, \citenamefont {Pouget-Abadie}, \citenamefont {Mirza},
  \citenamefont {Xu}, \citenamefont {Warde-Farley}, \citenamefont {Ozair},
  \citenamefont {Courville},\ and\ \citenamefont
  {Bengio}}]{goodfellow_generative_2014}%
  \BibitemOpen
  \bibfield  {author} {\bibinfo {author} {\bibfnamefont {I.}~\bibnamefont
  {Goodfellow}}, \bibinfo {author} {\bibfnamefont {J.}~\bibnamefont
  {Pouget-Abadie}}, \bibinfo {author} {\bibfnamefont {M.}~\bibnamefont
  {Mirza}}, \bibinfo {author} {\bibfnamefont {B.}~\bibnamefont {Xu}}, \bibinfo
  {author} {\bibfnamefont {D.}~\bibnamefont {Warde-Farley}}, \bibinfo {author}
  {\bibfnamefont {S.}~\bibnamefont {Ozair}}, \bibinfo {author} {\bibfnamefont
  {A.}~\bibnamefont {Courville}}, \ and\ \bibinfo {author} {\bibfnamefont
  {Y.}~\bibnamefont {Bengio}},\ }\bibfield  {title} {\enquote {\bibinfo {title}
  {Generative adversarial nets},}\ }in\ \href
  {https://proceedings.neurips.cc/paper_files/paper/2014/file/5ca3e9b122f61f8f06494c97b1afccf3-Paper.pdf}
  {\emph {\bibinfo {booktitle} {Advances in Neural Information Processing
  Systems}}},\ Vol.~\bibinfo {volume} {27},\ \bibinfo {editor} {edited by\
  \bibinfo {editor} {\bibfnamefont {Z.}~\bibnamefont {Ghahramani}}, \bibinfo
  {editor} {\bibfnamefont {M.}~\bibnamefont {Welling}}, \bibinfo {editor}
  {\bibfnamefont {C.}~\bibnamefont {Cortes}}, \bibinfo {editor} {\bibfnamefont
  {N.}~\bibnamefont {Lawrence}}, \ and\ \bibinfo {editor} {\bibfnamefont
  {K.~Q.}\ \bibnamefont {Weinberger}}}\ (\bibinfo  {publisher} {Curran
  Associates, Inc.})\BibitemShut {NoStop}%
\bibitem [{\citenamefont {Karras}\ \emph {et~al.}(2018)\citenamefont {Karras},
  \citenamefont {Aila}, \citenamefont {Laine},\ and\ \citenamefont
  {Lehtinen}}]{karras_progressive_2017}%
  \BibitemOpen
  \bibfield  {author} {\bibinfo {author} {\bibfnamefont {T.}~\bibnamefont
  {Karras}}, \bibinfo {author} {\bibfnamefont {T.}~\bibnamefont {Aila}},
  \bibinfo {author} {\bibfnamefont {S.}~\bibnamefont {Laine}}, \ and\ \bibinfo
  {author} {\bibfnamefont {J.}~\bibnamefont {Lehtinen}},\ }\bibfield  {title}
  {\enquote {\bibinfo {title} {Progressive growing of gans for improved
  quality, stability, and variation},}\ }in\ \href@noop {} {\emph {\bibinfo
  {booktitle} {International Conference on Learning Representations}}}\
  (\bibinfo {year} {2018})\BibitemShut {NoStop}%
\bibitem [{\citenamefont {Yu}\ \emph {et~al.}()\citenamefont {Yu},
  \citenamefont {Zhang}, \citenamefont {Wang},\ and\ \citenamefont
  {Yu}}]{yu_seqgan_2017}%
  \BibitemOpen
  \bibfield  {author} {\bibinfo {author} {\bibfnamefont {L.}~\bibnamefont
  {Yu}}, \bibinfo {author} {\bibfnamefont {W.}~\bibnamefont {Zhang}}, \bibinfo
  {author} {\bibfnamefont {J.}~\bibnamefont {Wang}}, \ and\ \bibinfo {author}
  {\bibfnamefont {Y.}~\bibnamefont {Yu}},\ }\bibfield  {title} {\enquote
  {\bibinfo {title} {{SeqGAN}: Sequence generative adversarial nets with policy
  gradient},}\ }\href {\doibase 10.1609/aaai.v31i1.10804} {\bibfield  {journal}
  {\bibinfo  {journal} {Proceedings of the {AAAI} Conference on Artificial
  Intelligence}\ }\textbf {\bibinfo {volume} {31}},\
  10.1609/aaai.v31i1.10804}\BibitemShut {NoStop}%
\bibitem [{\citenamefont {Yi}, \citenamefont {Walia},\ and\ \citenamefont
  {Babyn}()}]{yi_generative_2019}%
  \BibitemOpen
  \bibfield  {author} {\bibinfo {author} {\bibfnamefont {X.}~\bibnamefont
  {Yi}}, \bibinfo {author} {\bibfnamefont {E.}~\bibnamefont {Walia}}, \ and\
  \bibinfo {author} {\bibfnamefont {P.}~\bibnamefont {Babyn}},\ }\bibfield
  {title} {\enquote {\bibinfo {title} {Generative adversarial network in
  medical imaging: A review},}\ }\href {\doibase
  https://doi.org/10.1016/j.media.2019.101552} {\bibfield  {journal} {\bibinfo
  {journal} {Medical Image Analysis}\ }\textbf {\bibinfo {volume} {58}},\
  \bibinfo {pages} {101552}}\BibitemShut {NoStop}%
\bibitem [{\citenamefont {Carleo}\ \emph {et~al.}()\citenamefont {Carleo},
  \citenamefont {Cirac}, \citenamefont {Cranmer}, \citenamefont {Daudet},
  \citenamefont {Schuld}, \citenamefont {Tishby}, \citenamefont
  {Vogt-Maranto},\ and\ \citenamefont {Zdeborová}}]{carleo_machine_2019}%
  \BibitemOpen
  \bibfield  {author} {\bibinfo {author} {\bibfnamefont {G.}~\bibnamefont
  {Carleo}}, \bibinfo {author} {\bibfnamefont {I.}~\bibnamefont {Cirac}},
  \bibinfo {author} {\bibfnamefont {K.}~\bibnamefont {Cranmer}}, \bibinfo
  {author} {\bibfnamefont {L.}~\bibnamefont {Daudet}}, \bibinfo {author}
  {\bibfnamefont {M.}~\bibnamefont {Schuld}}, \bibinfo {author} {\bibfnamefont
  {N.}~\bibnamefont {Tishby}}, \bibinfo {author} {\bibfnamefont
  {L.}~\bibnamefont {Vogt-Maranto}}, \ and\ \bibinfo {author} {\bibfnamefont
  {L.}~\bibnamefont {Zdeborová}},\ }\bibfield  {title} {\enquote {\bibinfo
  {title} {Machine learning and the physical sciences},}\ }\href {\doibase
  10.1103/RevModPhys.91.045002} {\bibfield  {journal} {\bibinfo  {journal}
  {Rev. Mod. Phys.}\ }\textbf {\bibinfo {volume} {91}},\ \bibinfo {pages}
  {045002}}\BibitemShut {NoStop}%
\bibitem [{\citenamefont {Mueller}, \citenamefont {Hernandez},\ and\
  \citenamefont {Wang}()}]{mueller_machine_2020}%
  \BibitemOpen
  \bibfield  {author} {\bibinfo {author} {\bibfnamefont {T.}~\bibnamefont
  {Mueller}}, \bibinfo {author} {\bibfnamefont {A.}~\bibnamefont {Hernandez}},
  \ and\ \bibinfo {author} {\bibfnamefont {C.}~\bibnamefont {Wang}},\
  }\bibfield  {title} {\enquote {\bibinfo {title} {Machine learning for
  interatomic potential models},}\ }\href {\doibase 10.1063/1.5126336}
  {\bibfield  {journal} {\bibinfo  {journal} {The Journal of Chemical Physics}\
  }\textbf {\bibinfo {volume} {152}},\ \bibinfo {pages} {050902}},\ \bibinfo
  {note} {\_eprint: https://doi.org/10.1063/1.5126336}\BibitemShut {NoStop}%
\bibitem [{\citenamefont {Friederich}\ \emph {et~al.}()\citenamefont
  {Friederich}, \citenamefont {Häse}, \citenamefont {Proppe},\ and\
  \citenamefont {Aspuru-Guzik}}]{friederich_machine-learned_2021}%
  \BibitemOpen
  \bibfield  {author} {\bibinfo {author} {\bibfnamefont {P.}~\bibnamefont
  {Friederich}}, \bibinfo {author} {\bibfnamefont {F.}~\bibnamefont {Häse}},
  \bibinfo {author} {\bibfnamefont {J.}~\bibnamefont {Proppe}}, \ and\ \bibinfo
  {author} {\bibfnamefont {A.}~\bibnamefont {Aspuru-Guzik}},\ }\bibfield
  {title} {\enquote {\bibinfo {title} {Machine-learned potentials for
  next-generation matter simulations},}\ }\href {\doibase
  10.1038/s41563-020-0777-6} {\bibfield  {journal} {\bibinfo  {journal} {Nature
  Materials}\ }\textbf {\bibinfo {volume} {20}},\ \bibinfo {pages}
  {750--761}}\BibitemShut {NoStop}%
\bibitem [{\citenamefont {Chung}\ \emph {et~al.}()\citenamefont {Chung},
  \citenamefont {Freitas}, \citenamefont {Cheon},\ and\ \citenamefont
  {Reed}}]{chung_data-centric_2022}%
  \BibitemOpen
  \bibfield  {author} {\bibinfo {author} {\bibfnamefont {H.~W.}\ \bibnamefont
  {Chung}}, \bibinfo {author} {\bibfnamefont {R.}~\bibnamefont {Freitas}},
  \bibinfo {author} {\bibfnamefont {G.}~\bibnamefont {Cheon}}, \ and\ \bibinfo
  {author} {\bibfnamefont {E.~J.}\ \bibnamefont {Reed}},\ }\bibfield  {title}
  {\enquote {\bibinfo {title} {Data-centric framework for crystal structure
  identification in atomistic simulations using machine learning},}\ }\href
  {\doibase 10.1103/PhysRevMaterials.6.043801} {\bibfield  {journal} {\bibinfo
  {journal} {Phys. Rev. Mater.}\ }\textbf {\bibinfo {volume} {6}},\ \bibinfo
  {pages} {043801}}\BibitemShut {NoStop}%
\bibitem [{\citenamefont {Yang}\ \emph {et~al.}()\citenamefont {Yang},
  \citenamefont {Cao}, \citenamefont {Zhang}, \citenamefont {Fan},
  \citenamefont {Tang}, \citenamefont {Aberg}, \citenamefont {Sadigh},\ and\
  \citenamefont {Zhou}}]{yang_self-supervised_2021}%
  \BibitemOpen
  \bibfield  {author} {\bibinfo {author} {\bibfnamefont {K.}~\bibnamefont
  {Yang}}, \bibinfo {author} {\bibfnamefont {Y.}~\bibnamefont {Cao}}, \bibinfo
  {author} {\bibfnamefont {Y.}~\bibnamefont {Zhang}}, \bibinfo {author}
  {\bibfnamefont {S.}~\bibnamefont {Fan}}, \bibinfo {author} {\bibfnamefont
  {M.}~\bibnamefont {Tang}}, \bibinfo {author} {\bibfnamefont {D.}~\bibnamefont
  {Aberg}}, \bibinfo {author} {\bibfnamefont {B.}~\bibnamefont {Sadigh}}, \
  and\ \bibinfo {author} {\bibfnamefont {F.}~\bibnamefont {Zhou}},\ }\bibfield
  {title} {\enquote {\bibinfo {title} {Self-supervised learning and prediction
  of microstructure evolution with convolutional recurrent neural networks},}\
  }\href {\doibase https://doi.org/10.1016/j.patter.2021.100243} {\bibfield
  {journal} {\bibinfo  {journal} {Patterns}\ }\textbf {\bibinfo {volume} {2}},\
  \bibinfo {pages} {100243}}\BibitemShut {NoStop}%
\bibitem [{\citenamefont {Lanzoni}\ \emph {et~al.}()\citenamefont {Lanzoni},
  \citenamefont {Albani}, \citenamefont {Bergamaschini},\ and\ \citenamefont
  {Montalenti}}]{lanzoni_morphological_2022}%
  \BibitemOpen
  \bibfield  {author} {\bibinfo {author} {\bibfnamefont {D.}~\bibnamefont
  {Lanzoni}}, \bibinfo {author} {\bibfnamefont {M.}~\bibnamefont {Albani}},
  \bibinfo {author} {\bibfnamefont {R.}~\bibnamefont {Bergamaschini}}, \ and\
  \bibinfo {author} {\bibfnamefont {F.}~\bibnamefont {Montalenti}},\ }\bibfield
   {title} {\enquote {\bibinfo {title} {Morphological evolution via surface
  diffusion learned by convolutional, recurrent neural networks: Extrapolation
  and prediction uncertainty},}\ }\href {\doibase
  10.1103/PhysRevMaterials.6.103801} {\bibfield  {journal} {\bibinfo  {journal}
  {Phys. Rev. Mater.}\ }\textbf {\bibinfo {volume} {6}},\ \bibinfo {pages}
  {103801}}\BibitemShut {NoStop}%
\bibitem [{\citenamefont {Brophy}\ \emph {et~al.}()\citenamefont {Brophy},
  \citenamefont {Wang}, \citenamefont {She},\ and\ \citenamefont
  {Ward}}]{brophy_generative_2023}%
  \BibitemOpen
  \bibfield  {author} {\bibinfo {author} {\bibfnamefont {E.}~\bibnamefont
  {Brophy}}, \bibinfo {author} {\bibfnamefont {Z.}~\bibnamefont {Wang}},
  \bibinfo {author} {\bibfnamefont {Q.}~\bibnamefont {She}}, \ and\ \bibinfo
  {author} {\bibfnamefont {T.}~\bibnamefont {Ward}},\ }\bibfield  {title}
  {\enquote {\bibinfo {title} {Generative adversarial networks in time series:
  A systematic literature review},}\ }\href {\doibase 10.1145/3559540}
  {\bibfield  {journal} {\bibinfo  {journal} {{ACM} Comput. Surv.}\ }\textbf
  {\bibinfo {volume} {55}},\ 10.1145/3559540}\BibitemShut {NoStop}%
\bibitem [{\citenamefont {Yeo}, \citenamefont {Li},\ and\ \citenamefont
  {Gifford}()}]{yeo_generative_2022}%
  \BibitemOpen
  \bibfield  {author} {\bibinfo {author} {\bibfnamefont {K.}~\bibnamefont
  {Yeo}}, \bibinfo {author} {\bibfnamefont {Z.}~\bibnamefont {Li}}, \ and\
  \bibinfo {author} {\bibfnamefont {W.}~\bibnamefont {Gifford}},\ }\bibfield
  {title} {\enquote {\bibinfo {title} {Generative adversarial network for
  probabilistic forecast of random dynamical systems},}\ }\href {\doibase
  10.1137/21M1457448} {\ \textbf {\bibinfo {volume} {44}},\ \bibinfo {pages}
  {A2150--A2175}},\ \bibinfo {note} {place: {USA} Publisher: Society for
  Industrial and Applied Mathematics}\BibitemShut {NoStop}%
\bibitem [{\citenamefont {Stinis}, \citenamefont {Daskalakis},\ and\
  \citenamefont {Atzberger}()}]{stinis_sdyn-gans_2023}%
  \BibitemOpen
  \bibfield  {author} {\bibinfo {author} {\bibfnamefont {P.}~\bibnamefont
  {Stinis}}, \bibinfo {author} {\bibfnamefont {C.}~\bibnamefont {Daskalakis}},
  \ and\ \bibinfo {author} {\bibfnamefont {P.~J.}\ \bibnamefont {Atzberger}},\
  }\bibfield  {title} {\enquote {\bibinfo {title} {{SDYN}-{GANs}: Adversarial
  learning methods for multistep generative models for general order stochastic
  dynamics},}\ }\href {\doibase https://doi.org/10.48550/arXiv.2302.03663} {\
  https://doi.org/10.48550/arXiv.2302.03663},\ \bibinfo {note} {\_eprint:
  2302.03663}\BibitemShut {NoStop}%
\bibitem [{\citenamefont {Saxena}\ and\ \citenamefont
  {Cao}()}]{saxena_generative_2021}%
  \BibitemOpen
  \bibfield  {author} {\bibinfo {author} {\bibfnamefont {D.}~\bibnamefont
  {Saxena}}\ and\ \bibinfo {author} {\bibfnamefont {J.}~\bibnamefont {Cao}},\
  }\bibfield  {title} {\enquote {\bibinfo {title} {Generative adversarial
  networks ({GANs}): Challenges, solutions, and future directions},}\ }\href
  {\doibase 10.1145/3446374} {\bibfield  {journal} {\bibinfo  {journal} {{ACM}
  Comput. Surv.}\ }\textbf {\bibinfo {volume} {54}},\
  10.1145/3446374}\BibitemShut {NoStop}%
\bibitem [{\citenamefont {Radford}, \citenamefont {Metz},\ and\ \citenamefont
  {Chintala}()}]{radford_unsupervised_2015}%
  \BibitemOpen
  \bibfield  {author} {\bibinfo {author} {\bibfnamefont {A.}~\bibnamefont
  {Radford}}, \bibinfo {author} {\bibfnamefont {L.}~\bibnamefont {Metz}}, \
  and\ \bibinfo {author} {\bibfnamefont {S.}~\bibnamefont {Chintala}},\
  }\bibfield  {title} {\enquote {\bibinfo {title} {Unsupervised representation
  learning with deep convolutional generative adversarial networks},}\ }\href
  {https://arxiv.org/abs/1511.06434} {\ }\BibitemShut {NoStop}%
\bibitem [{\citenamefont {Goodfellow}()}]{goodfellow_nips_2017}%
  \BibitemOpen
  \bibfield  {author} {\bibinfo {author} {\bibfnamefont {I.}~\bibnamefont
  {Goodfellow}},\ }\bibfield  {title} {\enquote {\bibinfo {title} {{NIPS} 2016
  tutorial: Generative adversarial networks},}\ }\href
  {https://arxiv.org/abs/1701.00160} {\ }\bibinfo {note} {\_eprint:
  1701.00160}\BibitemShut {NoStop}%
\bibitem [{\citenamefont {Arjovsky}\ and\ \citenamefont
  {Bottou}()}]{arjovsky_towards_2017}%
  \BibitemOpen
  \bibfield  {author} {\bibinfo {author} {\bibfnamefont {M.}~\bibnamefont
  {Arjovsky}}\ and\ \bibinfo {author} {\bibfnamefont {L.}~\bibnamefont
  {Bottou}},\ }\bibfield  {title} {\enquote {\bibinfo {title} {Towards
  principled methods for training generative adversarial networks},}\ }\href
  {\doibase https://doi.org/10.48550/arXiv.1701.04862} {\
  https://doi.org/10.48550/arXiv.1701.04862},\ \bibinfo {note} {\_eprint:
  1701.04862}\BibitemShut {NoStop}%
\bibitem [{\citenamefont {Mescheder}, \citenamefont {Nowozin},\ and\
  \citenamefont {Geiger}()}]{mescheder_numerics_2017}%
  \BibitemOpen
  \bibfield  {author} {\bibinfo {author} {\bibfnamefont {L.}~\bibnamefont
  {Mescheder}}, \bibinfo {author} {\bibfnamefont {S.}~\bibnamefont {Nowozin}},
  \ and\ \bibinfo {author} {\bibfnamefont {A.}~\bibnamefont {Geiger}},\
  }\bibfield  {title} {\enquote {\bibinfo {title} {The numerics of {GANs}},}\
  }\href
  {https://proceedings.neurips.cc/paper_files/paper/2017/file/4588e674d3f0faf985047d4c3f13ed0d-Paper.pdf}
  {\ \textbf {\bibinfo {volume} {30}}}\BibitemShut {NoStop}%
\bibitem [{\citenamefont {Ho}, \citenamefont {Jain},\ and\ \citenamefont
  {Abbeel}()}]{ho_denoising_2020}%
  \BibitemOpen
  \bibfield  {author} {\bibinfo {author} {\bibfnamefont {J.}~\bibnamefont
  {Ho}}, \bibinfo {author} {\bibfnamefont {A.}~\bibnamefont {Jain}}, \ and\
  \bibinfo {author} {\bibfnamefont {P.}~\bibnamefont {Abbeel}},\ }\bibfield
  {title} {\enquote {\bibinfo {title} {Denoising diffusion probabilistic
  models},}\ }\href {https://arxiv.org/abs/2006.11239} {\ }\bibinfo {note}
  {\_eprint: 2006.11239}\BibitemShut {NoStop}%
\bibitem [{\citenamefont {Kingma}\ and\ \citenamefont
  {Welling}()}]{kingma_auto-encoding_2022}%
  \BibitemOpen
  \bibfield  {author} {\bibinfo {author} {\bibfnamefont {D.~P.}\ \bibnamefont
  {Kingma}}\ and\ \bibinfo {author} {\bibfnamefont {M.}~\bibnamefont
  {Welling}},\ }\bibfield  {title} {\enquote {\bibinfo {title} {Auto-encoding
  variational bayes},}\ }\href {https://arxiv.org/abs/1312.6114} {\ }\bibinfo
  {note} {\_eprint: 1312.6114}\BibitemShut {NoStop}%
\bibitem [{\citenamefont {Van~Kampen}()}]{van_kampen_stochastic_1992}%
  \BibitemOpen
  \bibfield  {author} {\bibinfo {author} {\bibfnamefont {N.~G.}\ \bibnamefont
  {Van~Kampen}},\ }\href@noop {} {\emph {\bibinfo {title} {Stochastic processes
  in physics and chemistry}}},\ Vol.~\bibinfo {volume} {1}\ (\bibinfo
  {publisher} {Elsevier})\BibitemShut {NoStop}%
\bibitem [{\citenamefont {Mirza}\ and\ \citenamefont
  {Osindero}(2014)}]{mirza_conditional_2014}%
  \BibitemOpen
  \bibfield  {author} {\bibinfo {author} {\bibfnamefont {M.}~\bibnamefont
  {Mirza}}\ and\ \bibinfo {author} {\bibfnamefont {S.}~\bibnamefont
  {Osindero}},\ }\bibfield  {title} {\enquote {\bibinfo {title} {Conditional
  generative adversarial nets},}\ }\href@noop {} {\bibfield  {journal}
  {\bibinfo  {journal} {arXiv preprint arXiv:1411.1784}\ } (\bibinfo {year}
  {2014})}\BibitemShut {NoStop}%
\bibitem [{\citenamefont {Goodfellow}, \citenamefont {Bengio},\ and\
  \citenamefont {Courville}()}]{goodfellow_deep_2016}%
  \BibitemOpen
  \bibfield  {author} {\bibinfo {author} {\bibfnamefont {I.}~\bibnamefont
  {Goodfellow}}, \bibinfo {author} {\bibfnamefont {Y.}~\bibnamefont {Bengio}},
  \ and\ \bibinfo {author} {\bibfnamefont {A.}~\bibnamefont {Courville}},\
  }\href@noop {} {\emph {\bibinfo {title} {Deep learning}}}\ (\bibinfo
  {publisher} {{MIT} press})\BibitemShut {NoStop}%
\bibitem [{\citenamefont {Paszke}\ \emph {et~al.}()\citenamefont {Paszke},
  \citenamefont {Gross}, \citenamefont {Massa}, \citenamefont {Lerer},
  \citenamefont {Bradbury}, \citenamefont {Chanan}, \citenamefont {Killeen},
  \citenamefont {Lin}, \citenamefont {Gimelshein}, \citenamefont {Antiga},
  \citenamefont {Desmaison}, \citenamefont {Kopf}, \citenamefont {Yang},
  \citenamefont {{DeVito}}, \citenamefont {Raison}, \citenamefont {Tejani},
  \citenamefont {Chilamkurthy}, \citenamefont {Steiner}, \citenamefont {Fang},
  \citenamefont {Bai},\ and\ \citenamefont {Chintala}}]{paszke_pytorch_2019}%
  \BibitemOpen
  \bibfield  {author} {\bibinfo {author} {\bibfnamefont {A.}~\bibnamefont
  {Paszke}}, \bibinfo {author} {\bibfnamefont {S.}~\bibnamefont {Gross}},
  \bibinfo {author} {\bibfnamefont {F.}~\bibnamefont {Massa}}, \bibinfo
  {author} {\bibfnamefont {A.}~\bibnamefont {Lerer}}, \bibinfo {author}
  {\bibfnamefont {J.}~\bibnamefont {Bradbury}}, \bibinfo {author}
  {\bibfnamefont {G.}~\bibnamefont {Chanan}}, \bibinfo {author} {\bibfnamefont
  {T.}~\bibnamefont {Killeen}}, \bibinfo {author} {\bibfnamefont
  {Z.}~\bibnamefont {Lin}}, \bibinfo {author} {\bibfnamefont {N.}~\bibnamefont
  {Gimelshein}}, \bibinfo {author} {\bibfnamefont {L.}~\bibnamefont {Antiga}},
  \bibinfo {author} {\bibfnamefont {A.}~\bibnamefont {Desmaison}}, \bibinfo
  {author} {\bibfnamefont {A.}~\bibnamefont {Kopf}}, \bibinfo {author}
  {\bibfnamefont {E.}~\bibnamefont {Yang}}, \bibinfo {author} {\bibfnamefont
  {Z.}~\bibnamefont {{DeVito}}}, \bibinfo {author} {\bibfnamefont
  {M.}~\bibnamefont {Raison}}, \bibinfo {author} {\bibfnamefont
  {A.}~\bibnamefont {Tejani}}, \bibinfo {author} {\bibfnamefont
  {S.}~\bibnamefont {Chilamkurthy}}, \bibinfo {author} {\bibfnamefont
  {B.}~\bibnamefont {Steiner}}, \bibinfo {author} {\bibfnamefont
  {L.}~\bibnamefont {Fang}}, \bibinfo {author} {\bibfnamefont {J.}~\bibnamefont
  {Bai}}, \ and\ \bibinfo {author} {\bibfnamefont {S.}~\bibnamefont
  {Chintala}},\ }\bibfield  {title} {\enquote {\bibinfo {title} {{PyTorch}: An
  imperative style, high-performance deep learning library},}\ }in\ \href
  {https://proceedings.neurips.cc/paper_files/paper/2019/file/bdbca288fee7f92f2bfa9f7012727740-Paper.pdf}
  {\emph {\bibinfo {booktitle} {Advances in Neural Information Processing
  Systems}}},\ Vol.~\bibinfo {volume} {32},\ \bibinfo {editor} {edited by\
  \bibinfo {editor} {\bibfnamefont {H.}~\bibnamefont {Wallach}}, \bibinfo
  {editor} {\bibfnamefont {H.}~\bibnamefont {Larochelle}}, \bibinfo {editor}
  {\bibfnamefont {A.}~\bibnamefont {Beygelzimer}}, \bibinfo {editor}
  {\bibfnamefont {F.~d.}\ \bibnamefont {Alché-Buc}}, \bibinfo {editor}
  {\bibfnamefont {E.}~\bibnamefont {Fox}}, \ and\ \bibinfo {editor}
  {\bibfnamefont {R.}~\bibnamefont {Garnett}}}\ (\bibinfo  {publisher} {Curran
  Associates, Inc.})\BibitemShut {NoStop}%
\bibitem [{\citenamefont {He}\ \emph {et~al.}(2016)\citenamefont {He},
  \citenamefont {Zhang}, \citenamefont {Ren},\ and\ \citenamefont
  {Sun}}]{he_deep_2015}%
  \BibitemOpen
  \bibfield  {author} {\bibinfo {author} {\bibfnamefont {K.}~\bibnamefont
  {He}}, \bibinfo {author} {\bibfnamefont {X.}~\bibnamefont {Zhang}}, \bibinfo
  {author} {\bibfnamefont {S.}~\bibnamefont {Ren}}, \ and\ \bibinfo {author}
  {\bibfnamefont {J.}~\bibnamefont {Sun}},\ }\bibfield  {title} {\enquote
  {\bibinfo {title} {Deep residual learning for image recognition},}\ }in\
  \href@noop {} {\emph {\bibinfo {booktitle} {Proceedings of the IEEE
  conference on computer vision and pattern recognition}}}\ (\bibinfo {year}
  {2016})\ pp.\ \bibinfo {pages} {770--778}\BibitemShut {NoStop}%
\bibitem [{\citenamefont {Kingma}\ and\ \citenamefont
  {Ba}()}]{kingma_adam_2014}%
  \BibitemOpen
  \bibfield  {author} {\bibinfo {author} {\bibfnamefont {D.~P.}\ \bibnamefont
  {Kingma}}\ and\ \bibinfo {author} {\bibfnamefont {J.}~\bibnamefont {Ba}},\
  }\bibfield  {title} {\enquote {\bibinfo {title} {Adam: A method for
  stochastic optimization},}\ }\href {\doibase
  https://doi.org/10.48550/arXiv.1412.6980} {\bibfield  {journal} {\bibinfo
  {journal} {{arXiv} preprint {arXiv}:1412.6980}\
  }https://doi.org/10.48550/arXiv.1412.6980}\BibitemShut {NoStop}%
\bibitem [{\citenamefont {Stinis}\ \emph {et~al.}()\citenamefont {Stinis},
  \citenamefont {Hagge}, \citenamefont {Tartakovsky},\ and\ \citenamefont
  {Yeung}}]{stinis_enforcing_2019}%
  \BibitemOpen
  \bibfield  {author} {\bibinfo {author} {\bibfnamefont {P.}~\bibnamefont
  {Stinis}}, \bibinfo {author} {\bibfnamefont {T.}~\bibnamefont {Hagge}},
  \bibinfo {author} {\bibfnamefont {A.~M.}\ \bibnamefont {Tartakovsky}}, \ and\
  \bibinfo {author} {\bibfnamefont {E.}~\bibnamefont {Yeung}},\ }\bibfield
  {title} {\enquote {\bibinfo {title} {Enforcing constraints for interpolation
  and extrapolation in generative adversarial networks},}\ }\href {\doibase
  10.1016/j.jcp.2019.07.042} {\bibfield  {journal} {\bibinfo  {journal}
  {Journal of Computational Physics}\ }\textbf {\bibinfo {volume} {397}},\
  \bibinfo {pages} {108844}}\BibitemShut {NoStop}%
\bibitem [{\citenamefont {Bishop}()}]{bishop_training_1995}%
  \BibitemOpen
  \bibfield  {author} {\bibinfo {author} {\bibfnamefont {C.~M.}\ \bibnamefont
  {Bishop}},\ }\bibfield  {title} {\enquote {\bibinfo {title} {Training with
  noise is equivalent to tikhonov regularization},}\ }\href {\doibase
  10.1162/neco.1995.7.1.108} {\bibfield  {journal} {\bibinfo  {journal} {Neural
  Computation}\ }\textbf {\bibinfo {volume} {7}},\ \bibinfo {pages}
  {108--116}}\BibitemShut {NoStop}%
\bibitem [{\citenamefont {Breiman}()}]{breiman_bagging_1996}%
  \BibitemOpen
  \bibfield  {author} {\bibinfo {author} {\bibfnamefont {L.}~\bibnamefont
  {Breiman}},\ }\bibfield  {title} {\enquote {\bibinfo {title} {Bagging
  predictors},}\ }\href {\doibase 10.1007/BF00058655} {\bibfield  {journal}
  {\bibinfo  {journal} {Machine Learning}\ }\textbf {\bibinfo {volume} {24}},\
  \bibinfo {pages} {123--140}}\BibitemShut {NoStop}%
\bibitem [{\citenamefont {Freund}\ and\ \citenamefont
  {Schapire}()}]{freund_desicion-theoretic_1995}%
  \BibitemOpen
  \bibfield  {author} {\bibinfo {author} {\bibfnamefont {Y.}~\bibnamefont
  {Freund}}\ and\ \bibinfo {author} {\bibfnamefont {R.~E.}\ \bibnamefont
  {Schapire}},\ }\bibfield  {title} {\enquote {\bibinfo {title} {A
  desicion-theoretic generalization of on-line learning and an application to
  boosting},}\ }\href {\doibase https://doi.org/10.1007/3-540-59119-2_166}
  {\bibinfo  {journal} {Computational Learning Theory}\ ,\ \bibinfo {pages}
  {23--37}}\BibitemShut {NoStop}%
\bibitem [{\citenamefont {Wolpert}()}]{wolpert_stacked_1992}%
  \BibitemOpen
\bibfield  {journal} {  }\bibfield  {author} {\bibinfo {author} {\bibfnamefont
  {D.~H.}\ \bibnamefont {Wolpert}},\ }\bibfield  {title} {\enquote {\bibinfo
  {title} {Stacked generalization},}\ }\href {\doibase
  https://doi.org/10.1016/S0893-6080(05)80023-1} {\bibfield  {journal}
  {\bibinfo  {journal} {Neural Networks}\ }\textbf {\bibinfo {volume} {5}},\
  \bibinfo {pages} {241--259}}\BibitemShut {NoStop}%
\bibitem [{\citenamefont {Ganaie}\ \emph {et~al.}()\citenamefont {Ganaie},
  \citenamefont {Hu}, \citenamefont {Malik}, \citenamefont {Tanveer},\ and\
  \citenamefont {Suganthan}}]{ganaie_ensemble_2022}%
  \BibitemOpen
  \bibfield  {author} {\bibinfo {author} {\bibfnamefont {M.~A.}\ \bibnamefont
  {Ganaie}}, \bibinfo {author} {\bibfnamefont {M.}~\bibnamefont {Hu}}, \bibinfo
  {author} {\bibfnamefont {A.~K.}\ \bibnamefont {Malik}}, \bibinfo {author}
  {\bibfnamefont {M.}~\bibnamefont {Tanveer}}, \ and\ \bibinfo {author}
  {\bibfnamefont {P.~N.}\ \bibnamefont {Suganthan}},\ }\bibfield  {title}
  {\enquote {\bibinfo {title} {Ensemble deep learning: A review},}\ }\href
  {\doibase https://doi.org/10.1016/j.engappai.2022.105151} {\bibfield
  {journal} {\bibinfo  {journal} {Engineering Applications of Artificial
  Intelligence}\ }\textbf {\bibinfo {volume} {115}},\ \bibinfo {pages}
  {105151}}\BibitemShut {NoStop}%
\bibitem [{\citenamefont {Tramèr}\ \emph {et~al.}()\citenamefont {Tramèr},
  \citenamefont {Kurakin}, \citenamefont {Papernot}, \citenamefont
  {Goodfellow}, \citenamefont {Boneh},\ and\ \citenamefont
  {{McDaniel}}}]{tramer_ensemble_2020}%
  \BibitemOpen
  \bibfield  {author} {\bibinfo {author} {\bibfnamefont {F.}~\bibnamefont
  {Tramèr}}, \bibinfo {author} {\bibfnamefont {A.}~\bibnamefont {Kurakin}},
  \bibinfo {author} {\bibfnamefont {N.}~\bibnamefont {Papernot}}, \bibinfo
  {author} {\bibfnamefont {I.}~\bibnamefont {Goodfellow}}, \bibinfo {author}
  {\bibfnamefont {D.}~\bibnamefont {Boneh}}, \ and\ \bibinfo {author}
  {\bibfnamefont {P.}~\bibnamefont {{McDaniel}}},\ }\bibfield  {title}
  {\enquote {\bibinfo {title} {Ensemble adversarial training: Attacks and
  defenses},}\ }\href {\doibase \_eprint: 1705.07204} {\ \_eprint:
  1705.07204},\ \bibinfo {note} {\_eprint: 1705.07204}\BibitemShut {NoStop}%
\bibitem [{\citenamefont {Ash}()}]{ash_basic_1970}%
  \BibitemOpen
  \bibfield  {author} {\bibinfo {author} {\bibfnamefont {R.~B.}\ \bibnamefont
  {Ash}},\ }\href@noop {} {\emph {\bibinfo {title} {Basic probability
  theory}}}\ (\bibinfo  {publisher} {John Wiley \& Sons, Inc.})\BibitemShut
  {NoStop}%
\bibitem [{\citenamefont {Dong}\ \emph {et~al.}(2020)\citenamefont {Dong},
  \citenamefont {Yu}, \citenamefont {Cao}, \citenamefont {Shi},\ and\
  \citenamefont {Ma}}]{Dong2020}%
  \BibitemOpen
  \bibfield  {author} {\bibinfo {author} {\bibfnamefont {X.}~\bibnamefont
  {Dong}}, \bibinfo {author} {\bibfnamefont {Z.}~\bibnamefont {Yu}}, \bibinfo
  {author} {\bibfnamefont {W.}~\bibnamefont {Cao}}, \bibinfo {author}
  {\bibfnamefont {Y.}~\bibnamefont {Shi}}, \ and\ \bibinfo {author}
  {\bibfnamefont {Q.}~\bibnamefont {Ma}},\ }\bibfield  {title} {\enquote
  {\bibinfo {title} {A survey on ensemble learning},}\ }\href@noop {}
  {\bibfield  {journal} {\bibinfo  {journal} {Frontiers of Computer Science}\
  }\textbf {\bibinfo {volume} {14}},\ \bibinfo {pages} {241--258} (\bibinfo
  {year} {2020})}\BibitemShut {NoStop}%
\bibitem [{\citenamefont {Zhou}(2021)}]{Zhou2021}%
  \BibitemOpen
  \bibfield  {author} {\bibinfo {author} {\bibfnamefont {Z.-H.}\ \bibnamefont
  {Zhou}},\ }\bibfield  {title} {\enquote {\bibinfo {title} {Ensemble
  learning},}\ }in\ \href {\doibase 10.1007/978-981-15-1967-3_8} {\emph
  {\bibinfo {booktitle} {Machine Learning}}}\ (\bibinfo  {publisher} {Springer
  Singapore},\ \bibinfo {address} {Singapore},\ \bibinfo {year} {2021})\ pp.\
  \bibinfo {pages} {181--210}\BibitemShut {NoStop}%
\bibitem [{\citenamefont {Esteban}, \citenamefont {Hyland},\ and\ \citenamefont
  {Rätsch}()}]{esteban_real-valued_2017}%
  \BibitemOpen
  \bibfield  {author} {\bibinfo {author} {\bibfnamefont {C.}~\bibnamefont
  {Esteban}}, \bibinfo {author} {\bibfnamefont {S.~L.}\ \bibnamefont {Hyland}},
  \ and\ \bibinfo {author} {\bibfnamefont {G.}~\bibnamefont {Rätsch}},\
  }\bibfield  {title} {\enquote {\bibinfo {title} {Real-valued (medical) time
  series generation with recurrent conditional {GANs}},}\ }\href {\doibase
  https://doi.org/10.48550/arXiv.1706.02633} {\
  https://doi.org/10.48550/arXiv.1706.02633},\ \bibinfo {note} {\_eprint:
  1706.02633}\BibitemShut {NoStop}%
\bibitem [{\citenamefont {Mogren}()}]{mogren_c-rnn-gan_2016}%
  \BibitemOpen
  \bibfield  {author} {\bibinfo {author} {\bibfnamefont {O.}~\bibnamefont
  {Mogren}},\ }\bibfield  {title} {\enquote {\bibinfo {title} {C-{RNN}-{GAN}:
  Continuous recurrent neural networks with adversarial training},}\ }\href
  {\doibase https://doi.org/10.48550/arXiv.1611.09904} {\
  https://doi.org/10.48550/arXiv.1611.09904},\ \bibinfo {note} {\_eprint:
  1611.09904}\BibitemShut {NoStop}%
\end{thebibliography}%

\end{document}